\def\simlt{\lower.5ex\hbox{$\; \buildrel < \over \sim \;$}}
\def\simgt{\lower.5ex\hbox{$\; \buildrel > \over \sim \;$}}
\RequirePackage{lineno} 

\documentclass[11pt,preprint]{aastex}

\newcommand{\myemail}{Mark.R.Swain@jpl.nasa.gov}
\slugcomment{Submitted to the Astrophysical Journal }
\shorttitle{Comparing HST/WFC3 stare and spatial scan observations}
\shortauthors{Swain et al.}

\begin{document}

\title{Comparing the WFC3 IR Grism Stare and Spatial-scan Observations 
for Exoplanet Characterization}

\author{Mark R. Swain}
\affil{Jet Propulsion Laboratory, California Institute of Technology, Pasadena, CA 91109}
\author{Pieter Deroo}
\affil{Jet Propulsion Laboratory, California Institute of Technology, Pasadena, CA 91109}
\author{Kiri L. Wagstaff}
\affil{Jet Propulsion Laboratory, California Institute of Technology, Pasadena, CA 91109}

\email{Mark.R.Swain@jpl.nasa.gov}
\altaffiltext{1}{Correspondence to be directed to \myemail}

\begin{abstract}

We report on a detailed study of the measurement stability for WFC3 IR
grism stare and spatial scan observations.  The excess measurement
noise for both modes is established by comparing the observed and
theoretical measurement uncertainties.  We find that the stare-mode
observations produce differential measurements that are consistent and
achieve $\sim1.3$ times photon-limited measurement precision.  In
contrast, the spatial-scan mode observations produce measurements
which are inconsistent, non-Gaussian, and have higher excess noise
corresponding to $\sim2$ times the photon-limited precision. The
inferior quality of the spatial scan observations is due to
spatial-temporal variability in the detector performance which we
measure and map. The non-Gaussian nature of spatial scan measurements
makes the use of $\chi^2$ and the determination of formal confidence
intervals problematic and thus renders the comparison of spatial scan
data with theoretical models or other measurements difficult. With
better measurement stability and no evidence for non-Gaussianity,
stare mode observations offer a significant advantage for
characterizing transiting exoplanet systems.

\end{abstract}

\keywords{methods: data analysis--instruments: individual(WFC3)}

\section{Introduction}

The objective of this paper is to characterize the measurement
stability of stare and spatial scan mode observations that have been
used to study exoplanets with WFC3. Because the spatial scan mode moves the
spectrum over a large range of detector pixels, which potentially have
different properties, it is important to test the measurement
stability as a function of location on the array. Exoplanet
spectroscopy via the transit method typically relies on taking the
difference of two large numbers to measure a small number.  This
approach makes knowledge of measurement systematic errors critically
important.  For characterizing an exoplanet's terminator transmission
spectrum (using the primary eclipse event) or dayside emission
spectrum (using the secondary eclipse event), the typical
spectroscopic measurement dynamic range varies from between
$\sim$1000:1 (1000 ppm) in the mid-infrared to $\sim$20,000:1 (50 ppm)
in the visible and near-infrared.  Continuing to improve the
measurement dynamic range is a priority.  Because higher precision
enables new science such as the detection of new opacity sources,
improved abundance constraints, ingress/egress light curve mapping,
measurement of atmospheric temporal variability, and characterization of
smaller targets, pushing for higher precision is critically important
to the field.  Most of the instruments used for transit spectroscopy
today were not specifically designed for the task; this makes
characterization of these instruments an especially important
activity.

Due to a combination of space-based location, good wavelength coverage,
moderate spectral resolution, and excellent telescope pointing, the
best instrument available today for transit spectroscopy is arguably
WFC3 using the IR grism.  The instrument has already been used
extensively for exoplanet spectroscopy (Berta et al. 2012, Swain et
al. 2013, Huitson et al. 2013, Wakeford et al. 2013, Deming, et
al. 2013, Mandell et al. 2013, Stevenson et al. 2013), and the field of
exoplanet atmospheric characterization is beginning to rely on this
critical community resource with programs conducted using both stare
and spatial scan observational modes.  Although a detailed comparison of
these modes did not previously exist, the instrument status report
(ISR) describing the spatial scan mode does note photometric `flicker'
(McCullough \& MacKenty 2012) at the 3\% level in spatial scan mode
testing which raises questions about how the mode will perform for
spectrophotometry measurements.  Motivated by the growing importance
of this instrument, we undertook a detailed analysis in which we
determined the measurement stability for differential spectroscopy,
the approach often used to characterize the atmospheres of transiting
planets.  WFC3 is an exceptional instrument that has produced
extraordinary measurements. Our findings, reported below, should be
viewed in the context of the astronomer's desire to operate an
instrument well beyond the original engineering design space.  Our
approach to characterizing the instrument is simple. Absent an
absolute calibration reference, the only way to assess instrument
stability is to make the same measurement many times and quantify the
repeatability.

\section{Methods}

\begin{figure*}
  \includegraphics[width=\hsize]{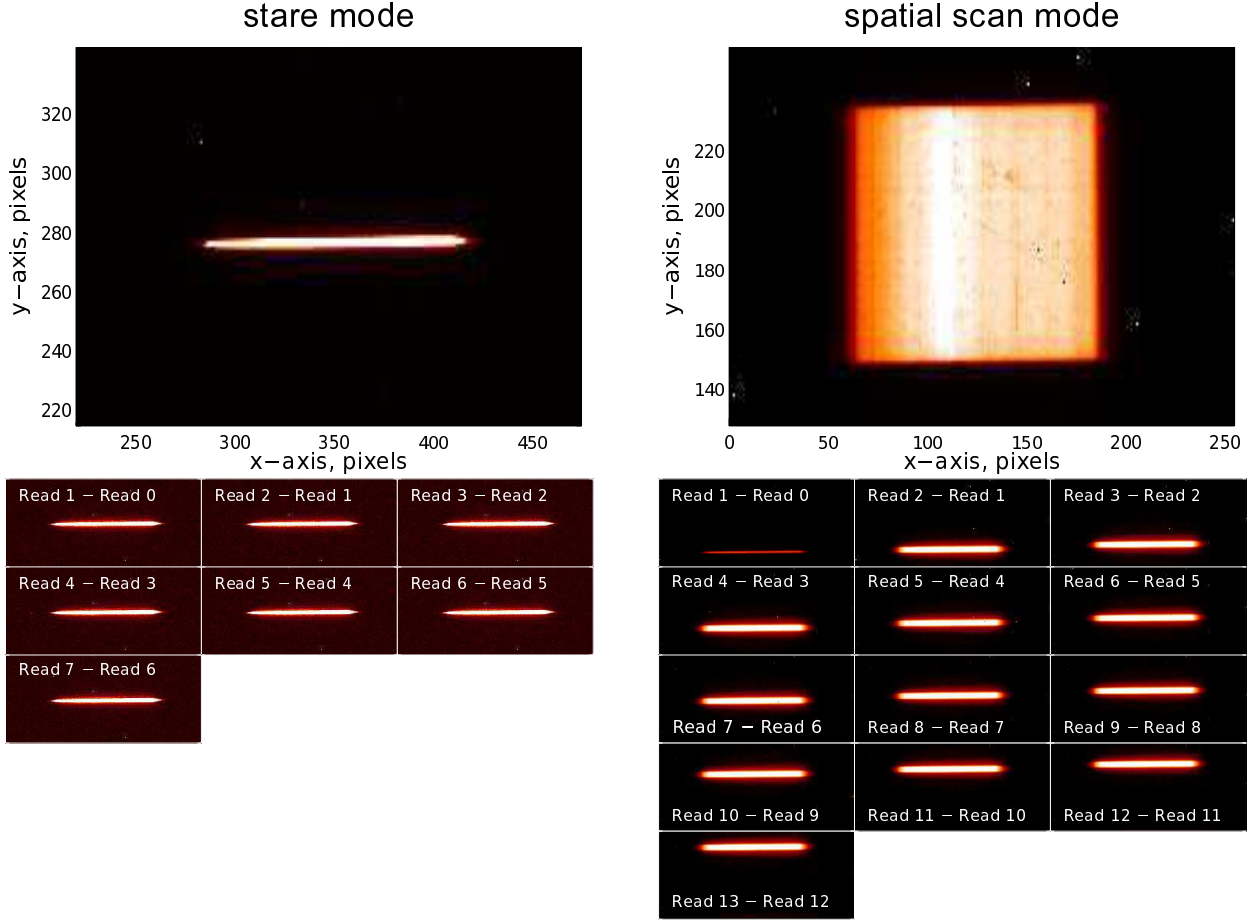} 
  \caption{\label{fig:figure1}An
  illustration of the difference between stare-mode observations
  (left), in which the spacecraft pointing is fixed in inertial space,
  and spatial-scan observations (right) in which the spacecraft is
  scanned in a direction perpendicular to the grism dispersion axis.
  The top figures are the respective images from a single integration
  consisting of a series of sample-up-the-ramp nondestructive reads.
  The lower figures show the difference between consecutive
  nondestructive reads. The stare observations have fewer
  nondestructive reads (Nsamp=8) than the spatial scan observations
  (Nsamp=14) shown here. The target of these observations is GJ
  1214. }
\end{figure*}


To quantify the stability of the WFC3 spectrophotometric time series
observations, we measure differences between portions of the time
series and determine how the observed noise compares to the
theoretically expected noise. Ideally, we would quantify the stability
using measurements of calibration targets with carefully selected
properties.  This could potentially require tens of Hubble orbits, and
no such calibration program has been carried out for WFC3 IR grism
spectroscopy.  An alternative approach is to base the analysis on a
science data set, preferably one where the same object has been
repeatedly observed in the same mode.  Fortunately, such a target
exists; the GJ 1214 exoplanet host star has been observed extensively
with the WFC3 IR grism using both stare and spatial scan mode
observations, and we use these data for our study. For all of the
observations we analyzed, the GJ 1214 system was observed with 4
consecutive HST orbits timed to place 2 orbits before the primary
eclipse event and one orbit after the primary eclipse event.  In our
analysis, the first orbit is excluded to allow for spacecraft settling
and detector effects that could uniquely affect the first orbit. We
analyze the difference between the pre- and post-eclipse time series
in which no exoplanet-related event impacts the measurement. The
differential measurement stability is characterized using the mean and
the variance.

All infrared observations with WFC3 are conducted in MULTIACCUM mode
which starts with two array resets, followed quickly by an initial
readout, and then a user selectable number of nondestructive reads,
specified by the MULTIACCUM mode and the Nsamp parameter (see WFC3
handbook).  The non-destructive reads (Nsamp=1 to Nsamp=N) sample up
the ramp as the detector accumulates electrons.  The initial readout
after a detector reset, termed the ''zero read'' (Nsamp=0), measures the
reset pedestal.  The light measured during a single integration is the
total number of photoelectrons measured during the sample up the ramp.
For WFC3, all detector reads, including the zero read, are saved and
available for download from the MAST archive. GJ 1214b data was
executed using both stare and scan observing modes (see Figure 1 and
Table 1) and in all cases, the detector was operated with the same
integration and number of nondestructive reads (Nsamp value) for each
of the four-orbit sequences.  In stare mode, the spectrum is held at
the same location and stabilized at the subpixel level on the detector
by the HST pointing system. The stare mode has frequently been used
for exoplanet spectroscopy because it minimizes sensitivity to inter
and intra-pixel non-uniformity.  Recently, the spatial scan mode has
been developed for WFC3 (McCullough \& MacKenty 2012) to improve
instrument duty cycle and bright target capability.  In this mode, the
spectrum is scanned across the detector in the spatial direction
during an integration; this process is then repeated in the same
detector location for subsequent integrations.

We measure instrument stability by constructing a quantity that should
be zero and tracking the behavior of this quantity.  Our merit figure for
instrument stability is the difference of the measured spectrum before
eclipse and after eclipse.  Subject to the assumption that the star is
constant, this difference should be zero; departures from zero
represent changes in the instrument spectral transfer function.  Any
deviations from zero, whatever the origin, represent aspects of the
measurement that must be accounted for when estimating the measurement
precision. To maintain sensitivity to array location-dependent
effects, we extract the electrons measured during each sample up the
ramp read.  We do this by using the $*$\_raw.fits data-type files, which
are the most minimally processed files available for WFC3.  

We begin the analysis by constructing a set of difference images,
$\Delta I$, which isolate the photoelectrons detected during a single
sample up the ramp read by taking the difference of consecutive
nondestructive reads,
\begin{equation}
\Delta I_{i} = I_{i} - I_{i-1},
\end{equation}
where $I$ is a single detector read image, and has time-related
indices corresponding to orbit, $o$, integration, $j$, and sample up
the ramp read, $i$.  This subtraction also removes array artifacts and
the background associated with previous reads. The $\Delta I_{i}$
images contain the spectrum + background measured during a specific
nondestructive read.  For each $\Delta I_{i}$ image (see Figure 1), we
determine the spectrum and background location, subtract the
background, and collapse the spectrum along the spatial axis. The
result is a spectrum, $F_{t}(\lambda)$, where $t$ can be decomposed
with indices representing orbit, $o$, integration number, $j$, and
sample up the ramp read number, $i$. To measure the achieved
stability, we take the coupled difference $\Delta F_{i,j}(\lambda)$ of
spectra measured in the pre and post eclipse Hubble orbit:
\begin{equation}
\Delta F_{i,j}(\lambda) = F_{pre,i,j}(\lambda) - F_{post,i,j}(\lambda),
\end{equation}
i.e. the first read of the first exposure of the pre-eclipse orbit is
subtracted from the first read of the first exposure of the post
eclipse orbit, etc. (see Figure 2). Taking the coupled difference has
the effect of removing the repeatable effects within an orbit, like a
ramp buildup.  We then construct an average quantity that removes the
dependence on integration, index $j$,
\begin{equation}
\Delta S_{i}(\lambda) = \frac{\langle \Delta F_{i,j}(\lambda) \rangle}
{\langle F_{pre,i,j}(\lambda), F_{post,i,j}(\lambda)\rangle}.
\end{equation}
The quantity $\Delta S_{i}(\lambda)$ measures the normalized
difference between spectra in orbits on either side of eclipse and is
a merit figure for spectral measurement stability. Ideally, it should
be zero; departures from zero indicate that the instrument spectral
transfer function has changed. For each of the $\Delta
S_{i}(\lambda)$, we determine the variance
\begin{equation}
\sigma^{2}_{\Delta S_{i}} = VAR \left[\frac{\Delta F_{i,j}(\lambda)}
{\langle F_{pre,i,j}(\lambda), F_{post,i,j}(\lambda)}\right]_{j},
\end{equation}
where VAR is the variance taken over the set of $j$ integrations. We
also calculate the average of the $\Delta S_{i}(\lambda)$, which
removes the dependence on the integration index, $i$,
\begin{equation}
\Delta S(\lambda) = \langle \Delta S_{i}(\lambda) \rangle,
\end{equation}
where the variance of $\Delta S(\lambda)$ is given by,
\begin{equation}
\sigma^{2}_{\Delta S} = VAR[\Delta S_{i}(\lambda)]_{i},
\end{equation}
and where VAR is the variance taken over the set of $i$
sample-up-the-ramp reads. We also calculate the photon noise for the
observations as,
\begin{equation}
\sigma^{2}_{phot} = N_{pre} + N_{post},
\end{equation}
where $N$ is the number of photons, and add additional noise for the
measurement noise due to the read process and the background.  By
comparing the photon noise (i.e.,~the theoretical limit) to the
measured noise, we determine the excess noise in the system.

For the spatial scan observations we analyzed, the reads are evenly
spaced in time after read 1.  Thus for the spatial-scan observations
with Nsamp=14 or 16, there are 14 or 16 nondestructive reads
(including the zero read), and in our analysis we use the difference
between reads $i$ and $i-1$ from read 2 through 13 or 15, which
corresponds to a set of 12 or 14 independent measurements, [$\Delta
S_{2}(\lambda)$...$\Delta S_{13} (\lambda)$], which can be used to
study measurement repeatability. The observations conducted in stare
mode have 8 samples up the ramp and the integration time is identical
starting from the zero read. For those observations, we construct 7
independent measurements. If the measurements are repeatable, $\Delta
S_{i}(\lambda) = \Delta S_{k}(\lambda)$, where $i$ and $k$ indicate
different sample-up-the-ramp reads.

\begin{deluxetable}{lrccccc}
\centering
\tablecolumns{7} 
\tablewidth{0pt}  
\tablecaption{\label{tab:table1} Data used in this analysis.}
\tablehead{ 
visit & \multicolumn{1}{c}{date} & 
mode & Nsamp & scan length & 
integrations & integration \\
 &
 &
 &
 &
\multicolumn{1}{c}{(pixels)} &
\multicolumn{1}{c}{per orbit} &
\multicolumn{1}{c}{time (s)} 
}
\startdata
stare visit 1 &  9-Oct-10 & RAPID  & 8  &  NA   & 48 & 5.971 \\ 
stare visit 2 & 28-Mar-11 & RAPID  & 8  &  NA   & 48 & 5.971 \\
stare visit 3 & 23-Jul-11 & RAPID  & 8  &  NA   & 48 & 5.971 \\
scan visit 1 & 27-Sep-12 & SPARS10 & 14 & 80.76 & 17 & 88.436 \\
scan visit 2 &  3-Oct-12 & SPARS10 & 14 & 80.72 & 17 & 88.436 \\
scan visit 3 & 11-Oct-12 & SPARS10 & 14 & 80.66 & 17 & 88.436 \\
scan visit 4 & 19-Oct-12 & SPARS10 & 14 & 80.94 & 17 & 88.436 \\
scan visit 5 & 29-Jan-13 & SPARS10 & 14 & 81.22 & 17 & 88.436 \\
scan visit 6 & 13-Mar-13 & SPARS10 & 16 & 104.45 & 19 & 103.129 \\
scan visit 7 & 15-Mar-13 & SPARS10 & 16 & 104.87 & 19 & 103.129 \\
scan visit 8 & 27-Mar-13 & SPARS10 & 16 & 104.76 & 19 & 103.129 \\
scan visit 9 &  4-Apr-13 & SPARS10 & 16 & 104.95 & 19 & 103.129 \\
scan visit 10 & 12-Apr-13 & SPARS10 & 16 & 104.66 & 19 & 103.129 \\
scan visit 11 &  1-May-13 & SPARS10 & 16 & 104.60 & 19 & 103.129 \\
scan visit 12 &  6-Jul-13 & SPARS10 & 16 & 104.48 & 19 & 103.129 \\
scan visit 13 &  4-Aug-13 & SPARS10 & 16 & 104.24 & 19 & 103.129 \\
scan visit 14 & 20-Aug-13 & SPARS10 & 16 & 104.38 & 19 & 103.129 \\
\enddata
\end{deluxetable}

\begin{figure*}
  \centering
    \includegraphics[width=\hsize]{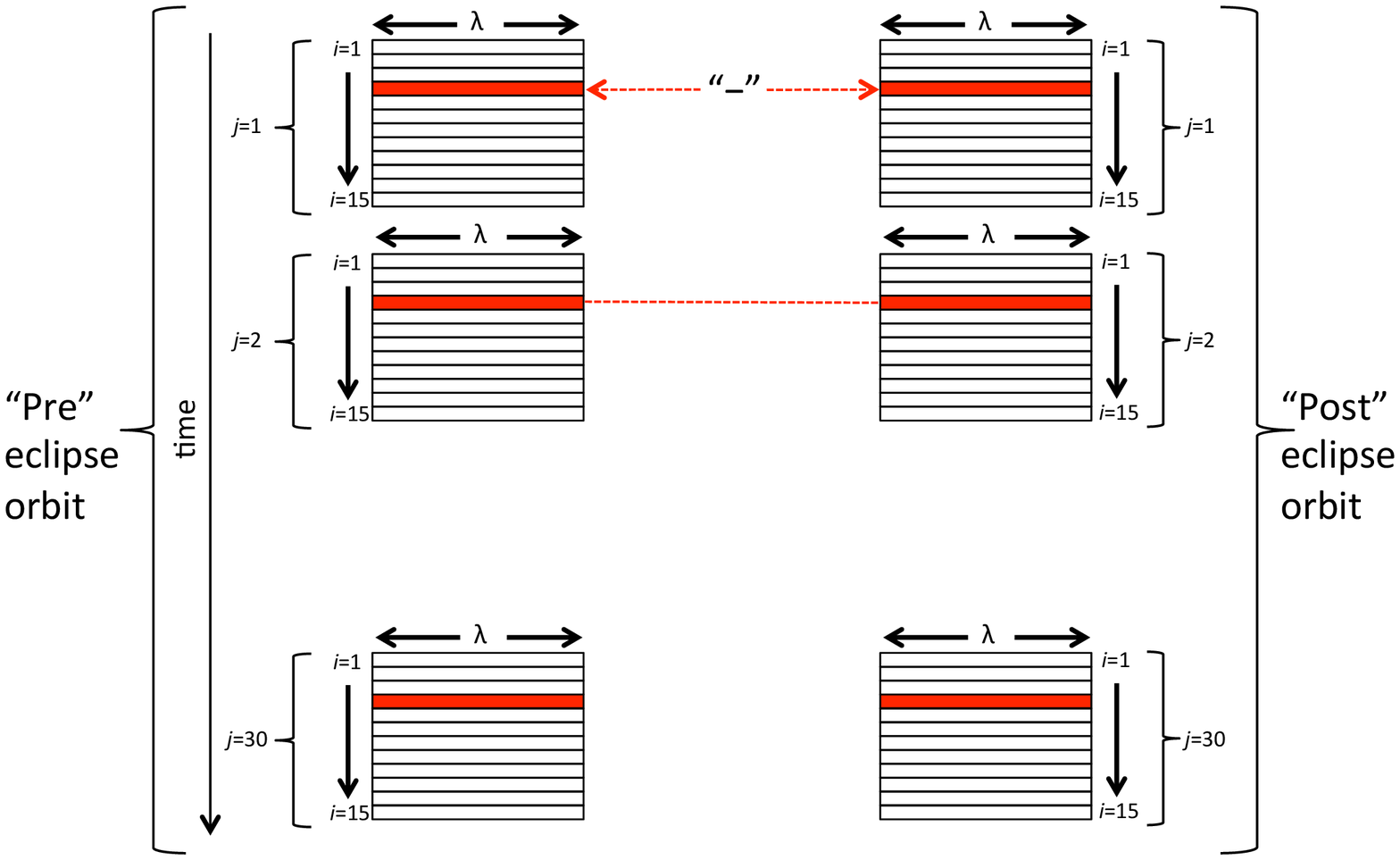}
     \caption{\label{fig:figure2}Diagram of data/index relationships illustrating how $\Delta S_{i}(\lambda)$ is constructed.  The red regions represent the data used in the equation above to construct $\Delta S_{4}(\lambda)$.}
\end{figure*}

\section{Results}

We investigated the excess noise and repeatability using the $\Delta
S_{i}$ and its statistics for a range of spectral resolutions by
averaging in the spectral direction.  These results are summarized for
single and multiple visits in Table 2 and Figure 3.  For the stare
mode, we find a relatively small amount of excess noise compared to
the photon noise and consistent $\Delta S_{i}$ measurements $(\Delta
S_{i} \approx \Delta S_{k})$. However, the spatial-scan mode
observations show significant excess noise and that $\Delta
S_{i} \neq \Delta S_{k}$ in the majority of cases, indicating the
measurements are not repeatable.  We tested the visit-to-visit
repeatability of the measurement by comparing the visit-averaged
$\Delta S$ and found that the repeatability is significantly better in
stare mode than for spatial scan mode, where we find significant
visit-to-visit variations (right panel of Figure 3).

\begin{center}
\begin{deluxetable}{ccc}
\tablecolumns{3} 
\tablewidth{0pt}  
\tablecaption{\label{tab:table2} Summary of findings for stare and spatial-scan modes.}
\tablehead{ \colhead{Observing mode} & \colhead{Condition} & \colhead{Comment} }
\startdata
 stare & $\sigma_{\Delta S_{i}} \approx \sigma_{photon}$  & Approaches theoretical noise limit $\sim 1.25 \times$ photon noise \\
       & $\Delta S_{i} \approx \Delta S_{k}$ & Measurements repeatable \\
       & $\Delta S_{i}$ Gaussian & interpretation straightforward \\
 spatial scan & $\delta_{\Delta S_{i}} > \delta_{photon}$ & Significant excess noise, up to more than $2\times$ photon noise \\
              & $\Delta S_{i} \neq \Delta S_{k}$ & Measurements not repeatable \\
              & $\Delta S_{i}$ non-Gaussian & science interpretation problematic \enddata
\end{deluxetable}
\end{center}	

\begin{figure*}
  \centering
    \includegraphics[width=\hsize]{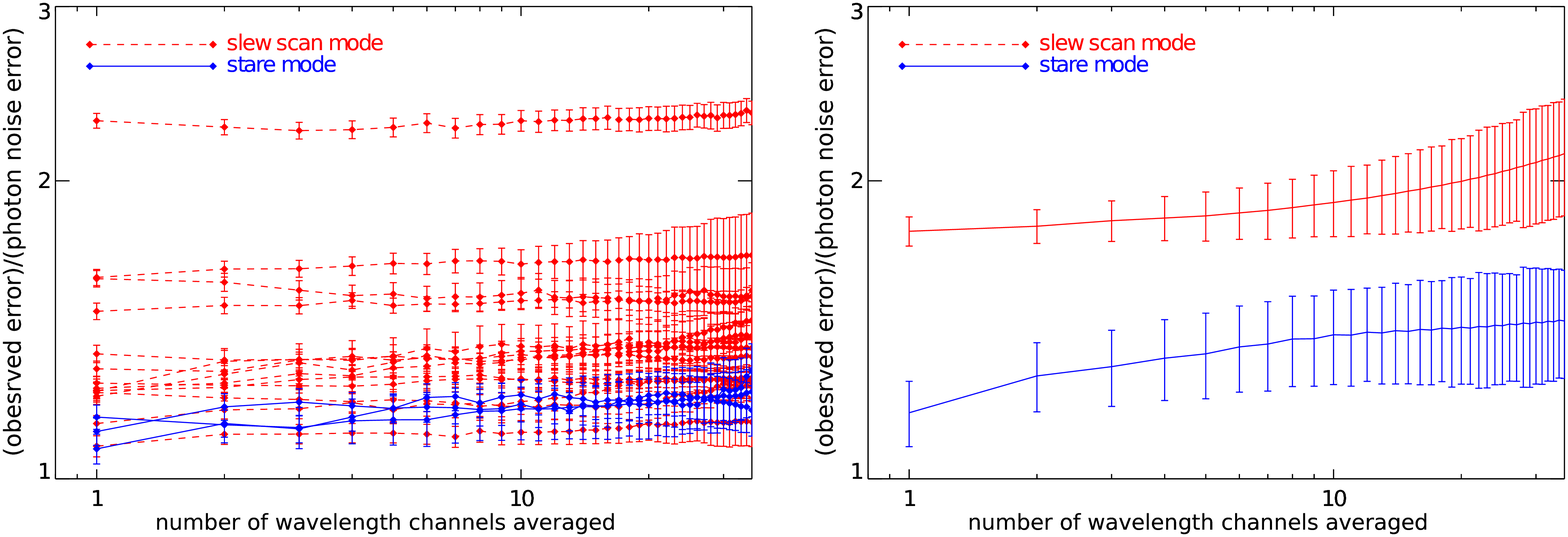}
     \caption{\label{fig:figure2}Excess noise for stare and spatial-scan modes as a function of spectral averaging for individual visits (left) and the multivisit average (right). The 1 $\sigma$ error bars indicate the differences observed when repeating the analysis for different wavelengths in the spectrum covering 1.15 to 1.65 $\mu$m. }
\end{figure*}

In addition to analyzing the variance of $\Delta S_{i}$, we also
analyzed its statistical properties. Given an unbiased source of
noise, we expect the $\Delta S_{i}$ measurements to follow a normal
distribution around the mean, $\Delta S$.  We analyzed the stare and
spatial scanning observations using a set of statistical tests,
including the $\chi^{2}$, Kolmogorov-Smirnov, Shapiro-Wilk,
Anderson-Darling, and D'Agostino-Pearson tests.  Each test generates a
statistic and an associated $p$-value.  While the range and meaning of
each test's calculated statistic varies, the $p$-value provides a
consistent way to interpret results across tests.  The $p$-value is the
probability of observing the given data values under the assumption
that the null hypothesis (here, of Gaussianity) is true.  A low
$p$-value provides evidence that the null hypothesis may not apply to
the observed data; e.g., a $p$-value of 0.04 indicates that the null
hypothesis can be rejected at the 95\% confidence level, because it is
less than 0.05.

The $\chi^{2}$ test compares two distributions represented as discrete
histograms, so the choice of histogram granularity can influence the
results.  We used 10 bins in all $\chi^{2}$ tests and compared the
observed frequencies to those expected from a random variable with a
Gaussian distribution having the same mean and standard deviation as
the observed data.

The tests were conducted on the measured $\Delta S_{i}$ spectra
for a given visit at 1.10 to 1.65 $\mu$m.  For the spatial-scan mode,
observations for all reads were combined.  We also tested the
collection of all observations across all visits for each mode (stare
and spatial scan).  The $p$-values for the stare mode data, across all
tests, are consistently too high for us to reject the Gaussian
hypothesis (see Table 3, ``all stare'').  In contrast, the Gaussian
hypothesis is strongly rejected for the spatial-scan mode data, by all
tests, with $>99$\% confidence (see Table 3, ``all scan'').  We also
examined the data from each visit individually.  For the stare mode,
the Gaussian hypothesis once again cannot be rejected with confidence
for any visit.  However, for the spatial-scan mode, the Gaussian
hypothesis is rejected with 99\% confidence for the majority of the
visits (2-6, 8, 9, 11, 13, and 14). ).  The remaining four visits (1,
7, 10, and 12) do not have sufficient evidence to reject the Gaussian
hypothesis with 99\% confidence, given the number of observations
available in each spatial 
scan visit.  Yet even in cases for which noise may be normally
distributed during a given visit, the noise is not consistent across visits,
implying that quantities estimated by averaging data from multiple
visits have an uncertainty that does not decrease as the square root
of the number of visits.

\begin{figure*}
  \centering
    \includegraphics[width=\hsize]{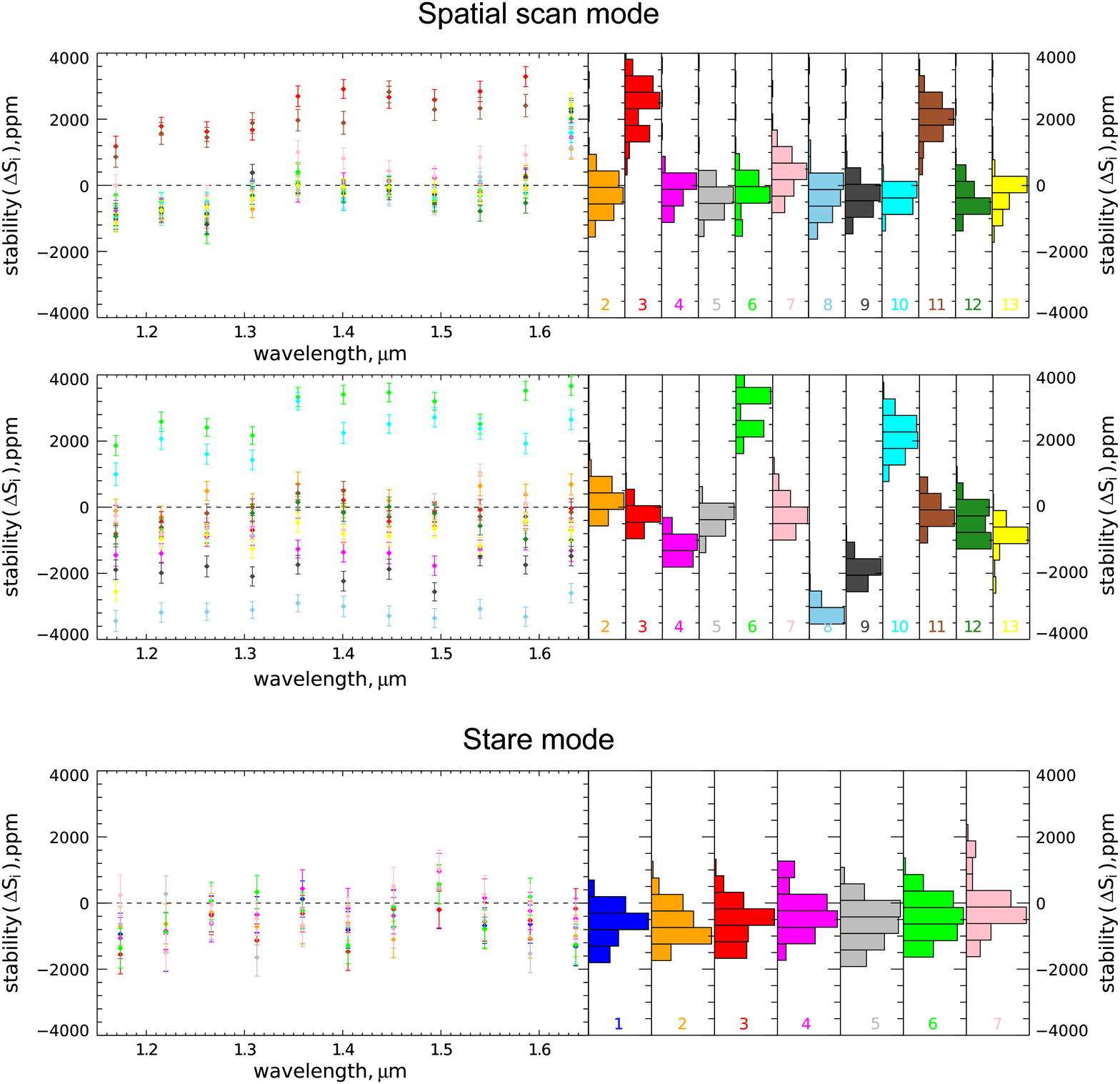}
     \caption{\label{fig:figure2}Three examples showing that observations are not repeatable in spatial scan mode (scan mode visit 4 top and visit 5 middle) whilst being repeatable in stare mode (bottom).  Shown to the left are the orbit averaged pre-post spectra for individual detector reads with the histograms for individual detector reads shown to the right; ideally, all measurements would be zero within the errors and the departures from zero here are due to systematic errors.   The average values of individual detector reads, corresponding to measurements made seconds apart, can easily disagree at the level of $\sim2000$ ppm (left) and by several standard deviations using the spatial-scan mode.  Stare observations (right) are much more repeatable. }
\end{figure*}

\begin{center}
\begin{deluxetable}{lrrrrr}
\tabletypesize{\footnotesize}
\tablecolumns{6} 
\tablewidth{0pt}  
\tablecaption{\label{tab:table3} P-values for Gaussianity tests applied to 
stare and spatial scan data.}
\tablehead{
visit & 
\multicolumn{1}{c}{$\chi^2$} & 
\multicolumn{1}{c}{Kolmogorov-} & 
\multicolumn{1}{c}{Sharipo-} & 
\multicolumn{1}{c}{Anderson-} & 
\multicolumn{1}{c}{D'Agostino-}  \\
\multicolumn{1}{l}{} & 
\multicolumn{1}{l}{} & 
\multicolumn{1}{c}{Smirnov} & 
\multicolumn{1}{c}{Wilk} & 
\multicolumn{1}{c}{Darling} & 
\multicolumn{1}{c}{Pearson} 
}
\startdata
stare visit 1 & 0.39 & 0.83 & 0.85 & $>0.15$ & 0.91 \\ 
stare visit 2 & 0.54 & 0.88 & 0.85 & $>0.15$ & 0.71 \\
stare visit 3 & 0.84 & 0.77 & 0.71 & $>0.15$ & 0.29 \\ \hline
\bf{all stare} & \bf{0.99} & \bf{0.98} & \bf{0.97} & \bf{$>$0.15} &
\bf{0.58} \\ \hline

scan visit 1 & 0.08  & 0.55 & 0.05  & $>0.05$ & 0.17 \\
scan visit 2 & 2e-15 & 5e-4 & 5e-10 & $<0.01$ & 3e-11 \\
scan visit 3 & 3e-4  & 0.05 & 4e-5  & $<0.01$ & 1e-3 \\
scan visit 4 & 6e-13 & 5e-5 & 6e-9  & $<0.01$ & 9e-8 \\
scan visit 5 & 9e-8  & 0.01 & 3e-5  & $<0.01$ & 0.02 \\
scan visit 6 & 7e-13 & 0.02 & 6e-7  & $<0.01$ & 7e-8 \\
scan visit 7 & 0.35  & 0.41 & 0.03  & $>0.05$ & 0.03 \\
scan visit 8 & 8e-11 & 0.10 & 1e-5  & $<0.01$ & 1e-4 \\
scan visit 9 & 0.01  & 0.23 & 2e-3  & $<0.01$ & 7e-4 \\
scan visit 10 & 0.31  & 0.88 & 0.35  & $>0.15$ & 0.89 \\
scan visit 11 & 2e-5  & 0.04 & 2e-5  & $<0.01$ & 5e-5 \\
scan visit 12 & 0.96  & 0.66 & 0.62  & $>0.15$ & 0.49 \\
scan visit 13 & 2e-8  & 0.06 & 2e-6  & $<0.01$ & 7e-8 \\
scan visit 14 & 6e-35 & 3e-4 & 2e-11 & $<0.01$ & 5e-12 \\ \hline
\bf{all scan} & \bf{0.00} & \bf{0.00} & \bf{2e-31} & \bf{$<$0.01} &
\bf{3e-55} 
\enddata
\end{deluxetable}
\end{center}	

\section{Discussion}

The non-Gaussian nature of the spatial-scan data makes estimation of
statistical significance problematic as normal Gaussian statistics
cannot be assumed.  The departures from Gaussianity become even
stronger when multiple spatial-scan visits are combined.  The finding
that the spatial scan measurement variance decreases more slowly than
$1/N$ (where $N$ is the number of measurements) indicates that the
individual measurements are not fully statistically independent.  This
combination of non-Gaussanity and a lack of statistical independence
has three important implications for measurements of exoplanets using
the spatial scan mode.

\begin{enumerate}

\item  {\bf Comparison of a spatial scan result with other measurements becomes problematic} because the confidence interval is no longer well defined.  We can measure the standard deviation, but without knowledge of the underlying distribution, we cannot know what the statistical significance of the result is.  This has direct relevance to the comparison of results obtained with WFC3 spatial-scan measurements of XO-1b (Deming et al. 2013) and previous stare-mode observations with NICMOS (Tinetti et al. 2010).  If the spatial-scan measurements of XO-1b exhibit non-Gaussian properties similar to those of the spatial-scan data sets we analyzed, the formal confidence of the statistical comparison reported by Deming et al. (2013) likely requires revision.

\item {\bf Increasing measurement precision by repeated spatial-scan observations is observationally expensive} because the measurement variance does not decrease down as $1/N$.  The spatial scan multivisit excess noise significantly reduces the efficiency for improved parameter constraint.

\item {\bf Increasing spectral coverage with repeated spatial-scan observations is not straightforward} for the same reasons that comparison of a spatial-scan result with other measurements is problematic.  Combining spatial-scan observations made with the WFC3 IR G102 and G141 grisms has to be undertaken with specific consideration to the excess noise and potential non-Gaussanity in each visit.

\end{enumerate}

The finding that the spatial scan mode suffers from a spatial-temporal
measurement instability of up to 2000 ppm in these observations (see
Figures 4 and 5) has a number of implications with relevance both to
future HST observations and to transit spectroscopy measurements with
other infrared instruments.

\begin{itemize}
  
\item {\bf Stare mode is superior:} Over the last decade, the conventional wisdom has been to execute exoplanet spectroscopy observations using a stare mode. Our analysis confirms that paradigm, with the stare mode to be substantially more stable than the spatial-scan mode.  It is reasonable to extrapolate these results to other infrared instruments and conclude that the stare mode is superior.  Multi-epoch measurements or ingress/egress light curve mapping are examples of science questions that place a premium on measurement stability and repeatability. Because the spatial-scan mode has non-Gaussian statistics, it complicates the derivation of scientific results based on this data. 

\item {\bf Analyze individual sample-up-the-ramp reads:} It is commonly assumed that the extent of systematics present in transiting exoplanet observations can be identified by analysis of time series residuals based on detector integrations (where averaging the samples up the ramp has been done).  This assumption is incorrect for the spatial-scanning mode.  There are indications from previous analysis (Swain et al. 2013) that the assumption is incorrect for some stare observations as well.  The analysis reported here shows that using individual samples up the ramp allows for detecting and characterizing excess noise. By analyzing time series made from individual samples up the ramp, the statistical properties of the data and systematics can be better established and accounted for in follow-on analysis.

\item {\bf Measurement repeatability:} In this analysis, we have constructed a metric to directly probe measurement stability for: 1) the time scale corresponding to the center-to-center separation of orbit 2 and orbit 4 (3.2 hours) and 2) the visit-to-visit repeatability.  These are the relevant time scales for exploring the impact of measurement stability (or lack thereof) on spectroscopic measurements of eclipse depth.  The 3.2 hour time scale investigates the accuracy of the single visit exoplanet spectrum while the visit-to-visit repeatability assesses the ability to average spectra taken at different times or determine exoplanet variability. Other time scales need to be explicitly probed for understanding the impact of measurement stability on ingress/egress mapping.

\item {\bf Specialized calibration for spatial scan mode:} The results here suggest that specialized calibration methods need to be developed for optimal measurement precision using the spatial-scan mode.  A simple approach could be rejecting detector reads corresponding to specific $\Delta S_{i}$ values.  This would somewhat reduce the number of usable photons, but it would likely render the observed instrument noise more Gaussian.  This approach could also be applied to a multivisit analysis. Another approach could treat the time series for each read as a separate measurement and apply some of the existing calibration methods to the time series individually. 

\item {\bf Implications for JWST spatial scan:} An extensive analysis, similar to the one presented here but with the scope enlarged to include a variety of time scales and detector operation modes, will be critical for JWST exoplanet science.  The combination of the need for instantaneous, broad spectral coverage, and JWST’s large collecting area will result in significant science pressure for observing in the spatial-scan mode.  The HgCdTe H2RG detectors used by JWST provide extended wavelength coverage to 5$\mu$m, well beyond the 1.7$\mu$m cutoff for the WFC3 HgCdTe H1RG detectors.  The longer wavelength devices may have different operating characteristics, possibly making it harder to achieve measurement stability.  Thus, both characterization of measurement stability and development of spatial scan calibration methods are likely to be crucial for maximizing exoplanet science with JWST.

\item {\bf Implications for multiobject measurements:} The presence of spatial-temporal changes that affect measurement stability in the WFC3-IR detector implies that this behavior is likely present in other infrared instruments and may set a limit of $\sim1000$ ppm on multi-object calibration approaches.  In the case of ground-based measurements, multi-object photometry and spectroscopy is a common approach for observing exoplanet primary and secondary eclipse events.  However, as Figures 4 and 5 show, the measurement stability changes depending on detector location.  Even for a space-grade detector like the one in WFC3, the spectral behavior of two different parts of the detector can easily change by 1000 ppm on the time scale of $\sim3$ hours; if the stellar placement is unlucky, the spatial-temporal detector errors can be worse.  The WFC3 H1RG detector is closely related to the larger H2RG.  Both the H1RG and H2RG series detectors are common in infrared instruments making the findings presented here potentially widely applicable.  Our results clearly show that spatial-temporal changes in detector performance may be an important factor limiting the performance of multiobject calibration approaches for exoplanet primary and secondary eclipse observations.

\item {\bf Large scale analysis has significant value:} This is the
  second paper produced by our group that is based on a uniform
  analysis of the individual sample-up-the-ramp detector reads in $>60$ orbits of WFC3 data.  In both this paper and the previous work (Swain et al. 2013), we find a range of instrument behaviors.  In some operating modes, there are significant visit-to-visit changes in instrument performance.  In both papers we identify systematic errors that were present in some data sets that could not be detected by working with data averaged to the detector integration time.  The uniform analysis of large amounts of data at the individual detector read level is critical for setting individual observations in a context and for elucidating the patterns of instrument behavior that should inform best practice guidelines.  The context for specific observations may be critical for interpreting a particular science measurement.  


\end{itemize}

WFC3 is a top-of-the-line instrument and constitutes one of the best
resources available for characterizing exoplanet atmospheres.  Our
findings highlight the necessity and urgency for the continued
detailed characterization of this instrument.  Ideally, we would like
to understand the origin of the spatial-temporal measurement
variations that give rise to the spatial scan excess noise and
non-Gaussianity.  Especially in the case of the spatial-scan
observations, there is a need to quantify measurement repeatability at
the level of individual sample-up-the-ramp reads; doing so reveals
excess noise for spatial-scan observations by investigating the
statistical properties of the individual samples, a method not
feasible using data that has been averaged to the detector integration
time scale.  In recently published results, it has been the convention
to analyze spatial scan data by combining data from individual
sample-up-the-ramp reads into a total number of electrons measured
during the detector integration time.  The excess spatial-scan noise
identified in this analysis has not been previously reported and could
constitute an unaccounted for noise source in WFC3 scan observational
results; incorporating the uncertainty identified here will likely
increase the error bars, potentially by as much as 2$\times$ in some
cases.  Fully understanding the origin of the excess noise present in
spatial-scan observations measurements is a study outside the scope of
this paper.  However, it is probable that with further investigation,
calibration methods could be developed that would permit spatial-scan
observations to consistently reach near-photon-limited measurement
performance.

\begin{center}
\begin{deluxetable}{ccccccc}
\tablecolumns{7} 
\tablewidth{0pt}  
\tablecaption{\label{tab:table4}Optimal observing configuration for stare-mode observations.}
\tablehead{ H-mag & Half-well & mode  & Nsamp & Integration & frames & efficiency \\ 
  & (s)& & & (s) & per orbit & }
\startdata
 8.0 &  1.3 & RAPID   & 4  &  1.1 & 101 &  4\% \\ 
 8.5 &  2.1 & RAPID   & 7  &  1.9 &  90 &  7\% \\
 9.0 &  3.3 & RAPID   & 12 &  3.3 &  69 &  9\% \\
 9.5 &  5.3 & RAPID   & 15 &  4.2 &  62 & 10\% \\
10.0 &  8.4 & SPARS10 & 2  &  7.6 & 100 & 29\% \\
10.5 & 13.2 & SPARS10 & 3  & 15.0 &  76 & 44\% \\
11.0 & 21.0 & SPARS25 & 2  & 22.6 &  65 & 57\% \\
11.5 & 33.3 & SPARS25 & 2  & 22.6 &  65 & 57\% \\
12.0 & 52.6 & SPARS25 & 3  & 45.0 &  43 & 74\% \\
12.5 & 83.3 & SPARS25 & 5  & 89.7 &  25 & 86\% \enddata
\end{deluxetable}
\end{center}	

Our analysis shows that the most stable measurements are obtained in
the stare mode and it is therefore useful to explore the optimal
observing parameters. To do so, we use the STSCI tools to determine
the photon gathering efficiency for targets with a range of
brightnesses in stare mode. We used the exposure time calculator to
determine optimum integration times and used the phase II planning
tool to calculate the entire overhead involved per orbit [we assumed a
target in the Kepler-field]. Because the overheads associated with the
spatial scan are not fully defined yet, we use the observed 55 \% $-$
75 \% efficiency for the GJ 1214.  The results are summarized in Table
4.  When optimally configured for maximum throughput and stability,
based on using the 256 $\times$ 256 subarray, which was found to be
more stable than the 512 $\times$ 512 subarray (Swain et al. 2013),
the stare mode is preferable for sources with an H band magnitude of
10.0 or fainter.  For brighter sources, the improved observational
efficiency of the spatial-scan mode offsets the added scan-noise
penalty although the problem of non-Gaussianity remains.


\begin{figure}
  \centering
    \includegraphics[width=0.45\textwidth]{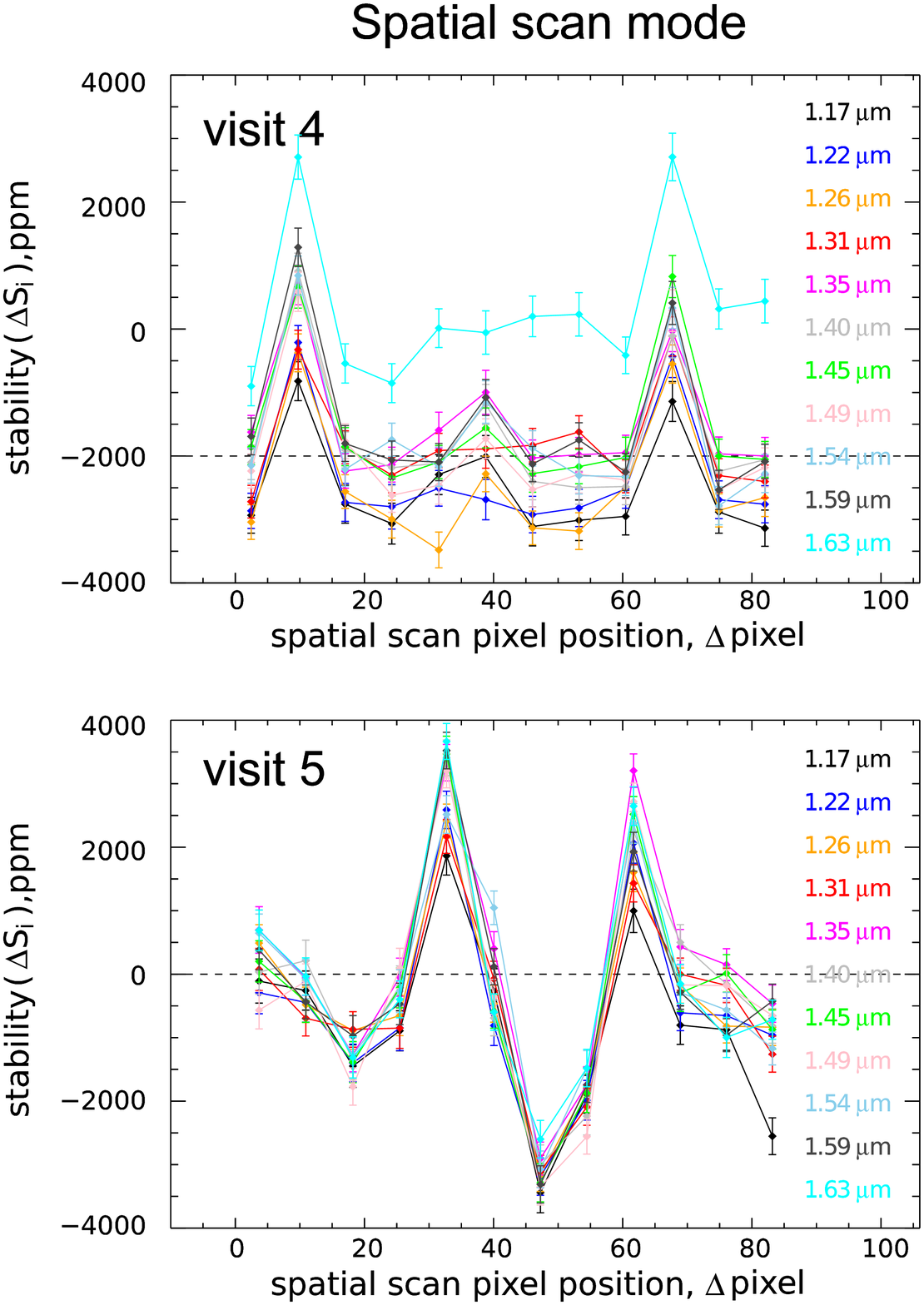}
     \caption{\label{fig:figure5}The spatial profile of the $\Delta S_{i}(\lambda)$ values for 11 spectral channels (spatial scan visit 4, see also Figure 3); ideally all values should be zero within the errors and departures from zero shown here represent systematic errors.  There is a clear dependence of the measurement property on location with measurement changes (nonrepeatability) of up to $\sim2000$ ppm.  The details of the spatial profiles change from visits to visit; in other words, the pixel-based location of measurement errors does not appear to be constant from visit to visit.}
\end{figure}

\section{Conclusions}


We compare the WFC3 stare and spatial scan observing modes in terms of
differential spectroscopy measurement performance and conclude that,
in terms of stability and repeatability, the stare mode is superior.
We show that the spatial scan observations exhibit non-Gaussian behavior
and suffer from significant excess measurement noise that is traceable
to changes in instrument behavior for different regions of the focal
plane.  This spatial-temporal variability creates excess noise at the
level of $\sim2 \times$ the photon noise that is relatively invariant
for changes in spectral resolution.  The spatial-temporal variations
correspond to a measurement error contribution of up to $\sim$75 ppm
per pixel per hour.  If other HAWAII-type detectors have similar
behavior, multiobject calibration approaches may be limited to
$\sim$1000 ppm measurement precision.  For JWST, which will use the
long wavelength H2RG detectors, careful investigation of the
spatial-scan noise and development of calibration methods is likely
crucial to achieve the full scientific potential of the observatory
for characterizing exoplanet atmospheres.

Only twice, this and one previous time, have large numbers of WFC3
orbits ($>60$) been systematically analyzed for measurement properties
at the level of individual sample-up-the-ramp detector reads.  In both
cases, evidence of systematics appeared that (1) could not be revealed
by the analysis of data averaged to the detector integration time, (2)
showed some instrument modes were better than others, and (3) found
that measurement performance in some instrument modes is variable at
the visit-to-visit time scale.  The approach of working with a time
series of individual detector sample-up-the-ramp reads allows for the
detection of systematic changes in eclipse depth or a stability
measurement metric that are correlated with detector read.  It also
provides a natural mechanism to handle saturated observations, which
occurs for a surprising number of the WFC3 exoplanet observations.

\section{Acknowledgements}
The authors thank Mark Colavita and Ingo Waldmann for comments on this
manuscript. The research described in this publication was carried out
in part at the Jet Propulsion Laboratory, California Institute of
Technology, under a contract with the National Aeronautics and Space
Administration. Copyright 2013 California Institute of
Technology. Government sponsorship acknowledged. All rights reserved.

\end{document}